\documentclass[aps,11pt,english,tightenlinesletterpaper,secnumarabic,nofootinbib,notitlepage]{revtex4-1}

\usepackage{graphicx}
\usepackage[justification=justified, singlelinecheck=off, compatibility=false]{caption}
\usepackage{subcaption}
\usepackage{amsmath}
\usepackage{bm}
\usepackage{amssymb}
\usepackage{color}
\usepackage{times}
\usepackage{float}
\usepackage{hyperref}
\hypersetup{
    colorlinks=true,       
    linkcolor=red,          
    citecolor=blue,        
    filecolor=magenta,      
    urlcolor=blue           
}

\newcommand{\eref}[1]{Eq.~(\ref{#1})}
\newcommand{\fref}[1]{Fig.~\ref{#1}}
\newcommand{\sref}[1]{Sec.~\ref{#1}}


\begin{document}

\title{Sensitivity of gravitational waves from preheating to a scalar field's interactions}

\date{\today}

\author{Jeffrey M. Hyde}
\email{jmhyde@asu.edu}
\affiliation{Physics Department, Arizona State University, Tempe, Arizona 85287, USA.}

\begin{abstract}
After inflation, a period of preheating may have produced a stochastic background of high frequency gravitational waves (GWs) that would persist until today. 
The nature of the inflaton's coupling to Standard Model or other fields is unknown, so it is useful to ask what features such fields may typically have, and how these affect predictions for the GW's produced. 
Here we consider the inflaton to be coupled to a light scalar field, and show that even a very small quartic self-interaction term will reduce the amplitude of the GW spectrum.
For self-coupling $\lambda_{\chi} \gtrsim g^2$, where $g^2$ is the inflaton-scalar coupling, the peak energy density goes as $\Omega_{\rm gw}^{(\lambda_{\chi})} / \Omega_{\rm gw}^{(\lambda_{\chi}=0)} \sim  (g^2/\lambda_{\chi})^{2}$. A consequence is that if the universe reheats through an inflaton-Higgs coupling then the spectrum would be suppressed but the dynamics would be sensitive to the Higgs potential near the energy scale of inflation.
\vspace{1cm}
\end{abstract}

\maketitle

\setlength{\parindent}{20pt}

\section{Introduction}\label{sec:intro}

Inflation leaves the universe cold and nearly empty of particles, so there needs to be a reheating mechanism for energy transfer between inflaton and Standard Model fields in order to create the thermalized particles that existed before Big Bang Nucleosynthesis began. This is typically modeled by a small, direct coupling between inflaton and another field. The first discussions of reheating \cite{Linde:1981mu,Albrecht:1982mp,Dolgov:1982th,Abbott:1982hn,Traschen:1990sw,Kofman:1994rk,Kofman:1997yn} studied a perturbative calculation of inflaton decay into the coupled field, with energy gradually transferred to matter fields. (Also see the reviews \cite{Allahverdi:2010xz,Amin:2014eta}.)

However, inflaton decay occurs in the context of large, coherent field oscillations and nonperturbative effects should also be taken into account \cite{Shtanov:1994ce,Kofman:1994rk,Kofman:1997yn,Greene:1997fu}. Typically, the inflaton $\phi$ is considered to be coupled to a field $\chi$ by an interaction $\frac{1}{2}g^2\phi^2\chi^2$, which is $\chi$'s only potential energy term. As the inflaton oscillates about the bottom of its potential after inflation, the phenomenon of parametric resonance leads to some modes of the decay product $\chi$ being excited at an exponential rate. This effect, which may occur briefly at the beginning of a longer period of reheating, is called preheating. (Most of the work on this subject has been in the context of direct couplings between inflaton and matter fields; see \cite{Alexander:2014bsa} for a scenario that does not require this.)

Preheating in these models can produce gravitational waves \cite{Khlebnikov:1997di,Easther:2006gt,Easther:2006vd,Dufaux:2007pt,GarciaBellido:2007af,Dufaux:2008dn}, since the exponential amplification of certain modes leads to a large contribution to anisotropic stress, which sources tensor perturbations. Predictions for the resulting spectrum are around $h^2\Omega_{\rm gw} \sim 10^{-10}$ and $f \sim 10^4$ to $10^6$ Hz today for massive or $\lambda\phi^4$ inflation or could be as low as $10^2$ to $10^3$ Hz for hybrid inflation models. Some work \cite{GarciaBellido:2007af,GarciaBellido:2008ab,Dufaux:2010cf,Enqvist:2012im,Enqvist:2012tc,Enqvist:2014tta,Figueroa:2014aya} has addressed this problem in the context of various models that relate to processes that are more specific. These find a wide range of possibilities. For example, \cite{Enqvist:2012im} found that decay into fermions after inflation could produce $\Omega_{\rm gw} \sim 10^{-12}$ to $10^{-18}$, $f \sim 10^{9}$ to $10^{10}$ Hz today, depending on the parameters.

These tend to fall outside the range of current, planned or proposed gravitational wave experiments such as Advanced LIGO and VIRGO, KAGRA, Einstein Telescope, eLISA, DECIGO or BBO (for an exception, see \cite{Dufaux:2010cf}). Roughly speaking, these are most sensitive to frequencies around $10^{-3}$ to $10^3$ Hz and signal strength corresponding to $h^2\Omega_{\rm gw} \sim 10^{-5}$ to possibly $10^{-14}$. (See \cite{Moore:2014lga,Thrane:2013oya} or the review \cite{Riles:2012yw}.\footnote{Note that some results are given in terms of $h^2\Omega_{\rm gw}$, others in terms of $\Omega_{\rm gw}$ and still others in terms of strain $h$, which is distinct from today's Hubble constant in units of 100 km/s/Mpc that appears in $h^2\Omega_{\rm gw}$. Consistent comparison of experimental sensitivities is discussed in \cite{Moore:2014lga}.}) LIGO and VIRGO have jointly placed upper bounds on a stochastic gravitational wave background on the order of $\Omega_{\rm gw} \sim 5 \times 10^{-6}$ around $10^2$ Hz \cite{Aasi:2014zwg}. Gravitational wave detection at MHz frequencies has also been considered \cite{Akutsu:2008qv,PhysRevD.77.022002,Goryachev:2014yra}. It has not been a major focus, though, since comparatively reliable astrophysical sources (e.g. neutron star mergers) are not expected in this frequency range.

This motivates the study of how robust are the predictions for the gravitational wave spectrum from preheating. We would expect that a realistic preheating process in the early universe would include couplings of the decay product to other fields, as well as possible self-interactions. It will be useful to know whether these can significantly affect the observability of such a process.\footnote{While this paper was in preparation, another work \cite{Lerner:2015uca} appeared that addresses some of these questions. We will discuss it in \sref{sec:discussion}.} Specifically, it would be interesting to answer the question ``Given a model of preheating with some self-interaction strength, how does one estimate the overall gravitational wave production?" This is analogous to the discussion in \cite{Giblin:2014dea}, which estimates the maximum energy density in gravitational waves that could be produced by a cosmological process such as preheating.

Previous work has shown that for self-couplings $\lambda_{\chi} \sim \mathcal{O}(10^{-2}) \gg g^2$, where $g^2$ is the coupling between the inflaton and scalar, parametric resonance can be significantly affected \cite{Prokopec:1996rr,Allahverdi:1996xc}. However, there has been little discussion of gravitational wave production in this scenario.\footnote{A study of gravitational waves in M-flation preheating \cite{Ashoorioon:2013oha} mentions that a self-interaction can suppress the resonance, but does not quantify this in a way that allows comparison with \cite{Prokopec:1996rr}.} Therefore it is difficult to give a thorough answer to the above question based on the existing literature. This also means that it is unclear how general a gravitational wave prediction is when it ignores interactions of the decay products.

In this work, we begin to address this by studying the development and termination of parametric resonance and the production of gravitational waves in the context of $\lambda\phi^4$ chaotic inflation coupled to a self-interacting light scalar field. We verify by lattice simulation that the resonance terminates early for self-coupling $\lambda_{\chi} \gtrsim g^2$, demonstrating the condition $\rho_{\chi}^{\rm final} \sim g^2/\lambda_{\chi}$ mentioned in footnote 19 of \cite{Prokopec:1996rr} (their $g$ is our $g^2$), and show that this leads to significant suppression of gravitational wave production. The resonance terminates early because the self-interaction term allows more efficient rescattering of particles out of the resonant mode, and this can be characterized by a condition comparing the energy density associated with the self-interaction to the inflaton-scalar interaction energy. The early termination of the resonance means that there is less energy in the light scalar's fluctuations, which directly source gravitational waves. Therefore, gravitational wave production is reduced. For $\lambda_{\chi} \gtrsim \lambda_{\chi}^{\ast} = g^2$, the energy density goes as $\Omega_{\rm gw}^{(\lambda_{\chi})} \sim (g^2/\lambda_{\chi})^2 \, \Omega_{\rm gw}^{(\lambda_{\chi}=0)}$.

In \sref{sec:generality} we show that this result is robust to changes in initial conditions, and that the same scaling occurs in massive ($m^2\phi^2$) inflation. Although this suggests generality to inflationary models that are quadratic or quartic about the minimum, we point out that an important goal of future work is to understand the effect of realistic interactions on other models that have predicted gravitational wave spectra.

As an application of this result, one could imagine the universe reheating by a coupling between the Higgs and inflaton, and we argue in \sref{sec:discussion} that such a scenario would likely produce no observable gravitational radiation. This is due to the size of the Higgs self-coupling, despite its eventual running to zero in the Standard Model. However, we point out that even a resonance too brief to produce observable gravitational waves could be relevant for the issue of vacuum stability. Finally, if the inflaton preheats a scalar field with an extremely small self-coupling, then the gravitational wave spectrum could directly measure this potential.

\section{Model}\label{sec:model}

Representing the universe by a spatially flat Friedmann-Robertson-Walker metric, we will describe gravitational waves as transverse and traceless perturbations to this metric, specifically as $h_{ij}$ such that 
\begin{align}\label{eq:metric}
	ds^2 = a^2(\eta)\left( -d\eta^2 + (\delta_{ij} + h_{ij})dx^idx^j \right)
\end{align}
with $\partial_ih_{ij} = 0$ and $h_{ii} = 0$. We will take the inflaton to be a real scalar field, $\phi(t,\vec{x})$, and consider it to be coupled to a massless real scalar field $\chi(t,\vec{x})$, with potential given by 
\begin{align}\label{eq:potential}
	V = \frac{1}{4}\lambda\phi^4 + \frac{1}{4}\lambda_{\chi}\chi^4 + \frac{1}{2}g^2\phi^2\chi^2
\end{align}
Here we have chosen to study $\lambda\phi^4$ chaotic inflation, and this requires some justification since standard slow-roll inflation with this potential is inconsistent with Cosmic Microwave Background (CMB) observations \cite{Ade:2015oja}. Much literature on gravitational waves from preheating takes the potential as $\frac{1}{4}\lambda\phi^4$, in particular the thorough numerical study \cite{Dufaux:2007pt}, whose model corresponds to ours with the choice $\lambda_{\chi}=0$. We expect the qualitative nature of our results to be relevant to a broad range of inflationary scenarios (this will be discussed further in \sref{sec:discussion}), and it will be useful to refer to specific previous results in order to understand the production of gravitational waves.

We are also studying the behavior of a ``light" scalar field, and so we neglect a $\chi$ mass term in comparison with the effective $\chi$ mass that comes from the interaction term $\frac{1}{2}g^2\phi^2\chi^2$. Comparing these terms using the amplitude of the $\phi$ oscillations shows that this is roughly equivalent to requiring the $\chi$ mass to be $m_{\chi} \ll \sqrt{g^2/\lambda} \times 10^{12}$ GeV.

Here the inflaton self-coupling is set by the amplitude of the scalar power spectrum of the CMB as $\lambda = 10^{-13}$. The unknown coupling $g^2$ must be small, but we will also take it to be larger than $\lambda$; in terms of the resonance parameter $q \equiv g^2/\lambda$ this means $1 \ll q \ll \lambda^{-1}$; here we will examine the range $10 \lesssim q \lesssim 2000$, which contains most of the region with the largest gravitational wave production \cite{Dufaux:2007pt}. We will see that this peaks around $q \approx 100-200$ and falls off slightly as $q$ gets larger or smaller (see \fref{fig:q_dependence}), although there are examples with smaller $q$ that do not exactly follow this trend \cite{Dufaux:2007pt}. We consider the light scalar's self-interaction in the range $\lambda < \lambda_{\chi} < 1$.

We study the dynamics in this model beginning at the end of inflation, $t_0 \equiv 0$, once the comoving horizon $(aH)^{-1}$ begins to expand, with the inflaton as a homogeneous field given everywhere by $\phi_0 = 0.342 \, M_{\rm Pl}$.\footnote{This particular point along the inflaton's phase space trajectory is identical to that of \cite{Dufaux:2007pt}. This choice is further addressed in \sref{sec:generality}.} The field $\chi$ is a light ``spectator" field during inflation, and at the end of inflation each $\chi$ mode is in the de Sitter vacuum state. As shown in previous work \cite{Khlebnikov:1996mc,Polarski:1995jg}, as the inflaton decays the quantum state quickly approaches a semiclassical regime with large occupation numbers, and the evolution here is equivalent to the classical evolution of an initial classical distribution that gives
\begin{align}
\langle |\chi_k(0)|^2\rangle = 1/(2\lambda^{3/2}\phi_0^3\omega_k), \, \, \, \, \, \, \, \, \dot{\chi}_k(0) = \left(i\omega_k + H(0)\right)\chi_k(0)
\end{align}
at the beginning of reheating.\footnote{$\chi_k$ and $\omega_k$ are defined below. The specific implementation for initial field conditions of \cite{Khlebnikov:1996mc} is as described in the documentation for LATTICEEASY code, available at http://www.felderbooks.com/latticeeasy/.} 
The dynamics considered here occurs on sub-horizon scales.\footnote{For the typical example $q=120$, numerical results show that preheating begins at about $H=1.1 \times 10^{-9} \, M_{\rm Pl}$ and $a=5.5$ (for $a=1$ at the beginning of the simulation) and the mode $k_{\ast} \approx \sqrt{\lambda}\phi_0$ is excited. Then at formation the wavelength of these perturbations is a fraction $R_{\ast} / R_{\rm horizon} = (a \, k_{\ast}^{-1}) H \sim 10^{-2}$ of the horizon size. Since inflation has ended, the comoving horizon $(aH)^{-1}$ is increasing, so $aH$ is decreasing and the modes excited later will be an even smaller fraction of the horizon size.}

Since $\phi$ is homogeneous, the equations of motion for these fields in a spatially flat Friedmann-Robertson-Walker (FRW) background are
\begin{align}
	\ddot{\phi} + 3H\dot{\phi} + \lambda\phi^3 & = 0 \label{eq:phi_eom} \\
	\square \chi + 3H\dot{\chi} + \lambda_{\chi}\chi^3 + g^2\phi^2\chi & = 0 \label{eq:chi_eom}
\end{align}
where $H \equiv \dot{a}/a$ is the Hubble parameter, whose value is related to the total energy density $\rho$ by the Friedmann equation
\begin{align}
	H^2 = \frac{8\pi G}{3} \rho.
\end{align}
In order to study the behavior of $\phi$ and $\chi$ that follows from the above, we will express $\chi$ in terms of modes $\chi_k$\footnote{Here we always use the Fourier Transform convention $f(\vec{x})=(2\pi)^{-3/2}\int d^3k \, f(\vec{k}) \, \exp(i\vec{k}\cdot\vec{x})$.}:
\begin{align}
	\chi(t,\vec{x}) = \frac{1}{(2\pi)^{3/2}}\int d^3k \left( a_k \chi_k(t) e^{-i\vec{k}\cdot\vec{x}} + a_k^{\dagger} \chi_k^{\ast}(t) e^{i\vec{k}\cdot\vec{x}} \right).
\end{align}
The amplitude of $\phi$ is still very large at the end of inflation, $\lambda_{\chi}\chi^2 \ll g^2\phi^2$, and \eref{eq:chi_eom} is approximately linear in $\chi$. We can then use the mode equation
\begin{align}\label{eq:linear_chi_eom}
	\ddot{\chi}_k + 3 H \dot{\chi}_k + \left( \frac{k^2}{a^2} + g^2\phi^2 \right)\chi_k = 0
\end{align}
to study the beginning of the reheating process. It will be convenient to introduce conformally rescaled fields $\overline{\phi} \equiv a\phi/\phi_0$, $\overline{\chi} \equiv a\chi/\phi_0$, and time $d\eta \equiv dt / a$ and define a dimensionless time parameter and wave number
\begin{align}\label{eq:x_def}
	\tau \equiv \sqrt{\lambda}\phi_0 \eta, \, \, \, \, \, \kappa \equiv k / (\sqrt{\lambda}\phi_0).
\end{align} 
Following e.g. \cite{Greene:1997fu,Dufaux:2007pt}, we study the field spectrum in terms of a comoving number density for the field $\chi$,
\begin{align}
	n_{\kappa} = \frac{1}{2}\left( \omega_{\kappa} |\overline{\chi}_{\kappa}|^2 + \frac{1}{\omega_{\kappa}}|\overline{\chi}_{\kappa}^{ \, \, '}|^2 \right),
\end{align}
and comoving energy density $\rho_{\kappa} = \omega_{\kappa} n_{\kappa}$, where $\omega_{\kappa} = \sqrt{\kappa^2 + m_{\rm eff}^2} = \sqrt{\kappa^2 + q\overline{\phi}^{\, 2} + 3(\lambda_{\chi}/\lambda)\overline{\chi}^{\, 2}}$.

\section{Preheating in This Model}\label{sec:preheating}

We begin by briefly outlining some results from previous studies of preheating, beginning with the case $\lambda_{\chi}=0$ (see \cite{Greene:1997fu} and references therein). We then use these to develop an approximate relation that quantifies the end of preheating and that will be useful in the gravitational wave calculation. After the end of inflation, $\phi$ oscillates in its potential with period $T \approx 7.416$ (in terms of the dimensionless time parameter $\tau$) \cite{Greene:1997fu} while the modes $\chi_{\kappa}$ can be excited by the phenomenon of parametric resonance. This process is typically described in terms of the resonance parameter $q = g^2/\lambda$. In general, certain modes $\kappa$ will be excited as $\chi_{\kappa} \propto \exp(\mu_{\kappa} \tau)$. The exponential growth factor $\mu_{\kappa}$ will vary with $\kappa$, giving rise to resonance ``bands" characterized by some central $\kappa$ and width $\Delta \kappa$. We will consider the case of ``broad resonance" where $q \gg 1$ (as compared with ``narrow resonance" when $q < 1$). In this case the spectrum of resonantly excited modes takes the form of a broad peak whose location and width are approximately characterized by 
\begin{align}
	\kappa_{\ast}, \, \Delta \kappa \, \, \, \sim \, \, \, q^{1/4}.
\end{align}
For a particular value of $q$, the maximum growth exponent $\mu_{\rm max} \equiv {\rm max}\{ \mu_{\kappa} \}$ is \cite{Greene:1997fu} 
\begin{align}
	\mu_{\rm max} & = \frac{1}{\pi} \ln \left( \sqrt{1 + \exp\left[-\pi \kappa^2\sqrt{2/q}\right]} + \exp\left[-\pi \kappa^2/\sqrt{2q}\right] \right)
\end{align}
and the resonance is efficient when $\kappa^2 \leq \sqrt{q/(2\pi^2)}$. Numerically we find that typical resonant momenta are $\kappa_{\ast}\sim 1$, so $\mu_{\rm max} \sim (3/2\pi)\exp(-\pi\sqrt{2/q})$ which is $\mathcal{O}(10^{-1})$ for the range of $q$ we consider. Number density $n_{\chi} \equiv \int d^3\kappa \, n_{\chi \, \kappa}$ increases in steps, twice per $\phi$ oscillation -- every time the inflaton passes through $\phi=0$ and $\chi$'s effective mass-squared $m_{\chi}^2 = g^2\phi^2$ goes to zero, a burst of $\chi$ particles are created. 

The exponential amplification of some $\chi_{\kappa}$ derived from \eref{eq:linear_chi_eom} is a solution for small $\chi$ (approximately zero) and homogeneous $\phi$, when the mode equation for $\chi_{\kappa}$ is linear. As this process evolves, this will become a worse approximation and the problem will become fully nonlinear. Therefore, \eref{eq:linear_chi_eom} is only useful for understanding the beginning of the reheating process, and in general it is the coupled equations of motion \eref{eq:phi_eom} and \eref{eq:chi_eom} that must be solved.


\begin{figure}[p]
\begin{center}
	\begin{subfigure}[b]{0.5\textwidth}
		\centering
		\includegraphics[width=\textwidth]{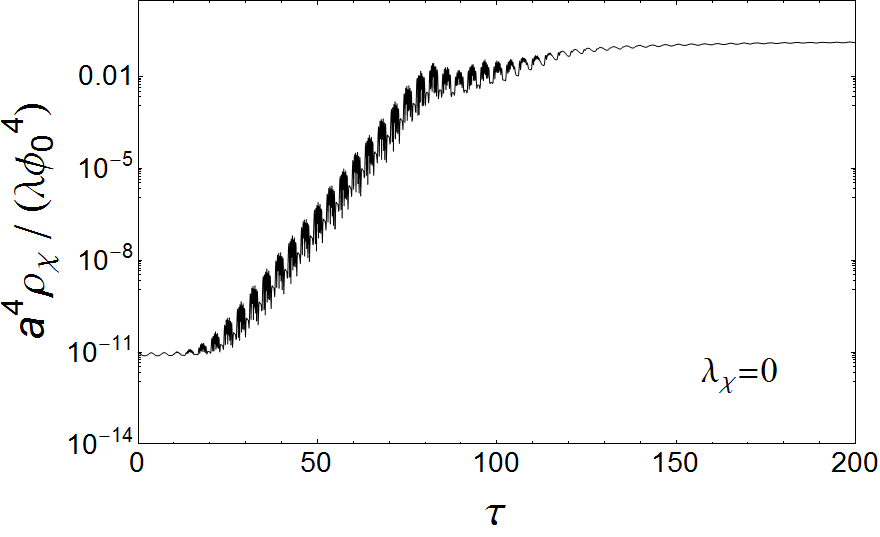}
                        \caption{\label{fig:egychi_120_0}}
	\end{subfigure}%
	\begin{subfigure}[b]{0.5\textwidth}
		\centering
		\includegraphics[width=\textwidth]{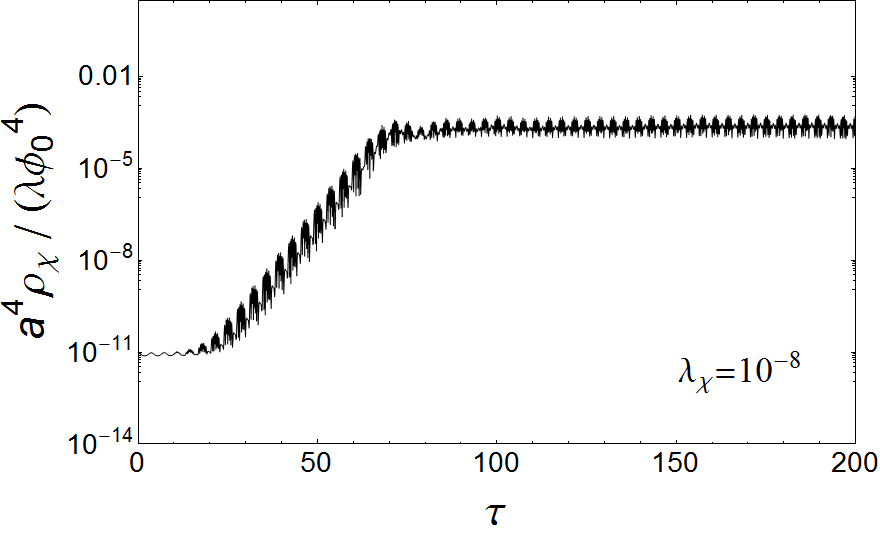}
                        \caption{\label{fig:egychi_120_8}}
	\end{subfigure}

	\begin{subfigure}[b]{0.5\textwidth}
		\centering
		\includegraphics[width=\textwidth]{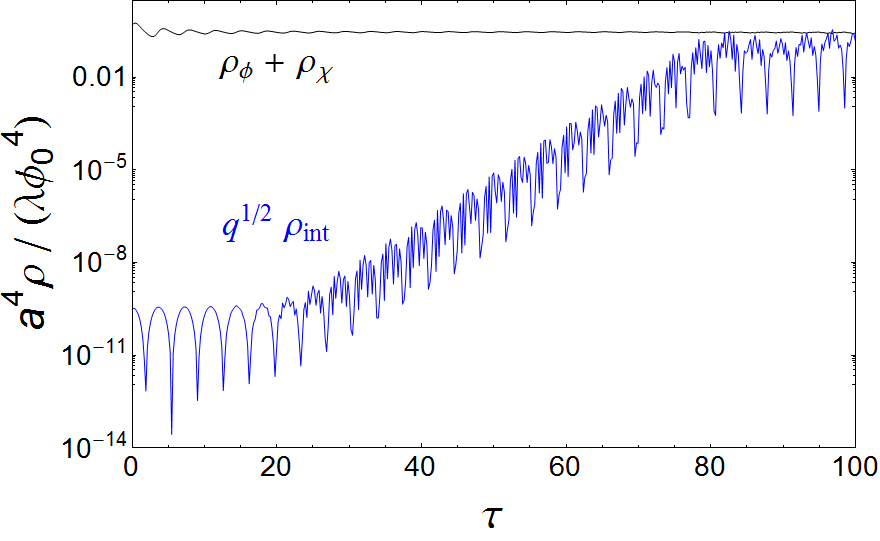}
                        \caption{\label{fig:egydens_120_0}}
	\end{subfigure}%
	\begin{subfigure}[b]{0.5\textwidth}
		\centering
		\includegraphics[width=\textwidth]{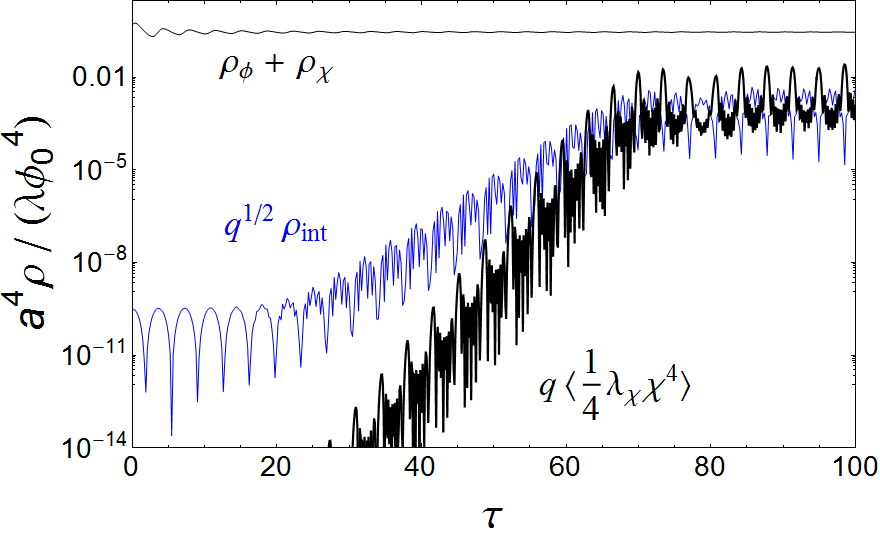}
                        \caption{\label{fig:egydens_120_8}}
	\end{subfigure}
	\begin{subfigure}[b]{0.5\textwidth}
		\centering
		\includegraphics[width=\textwidth]{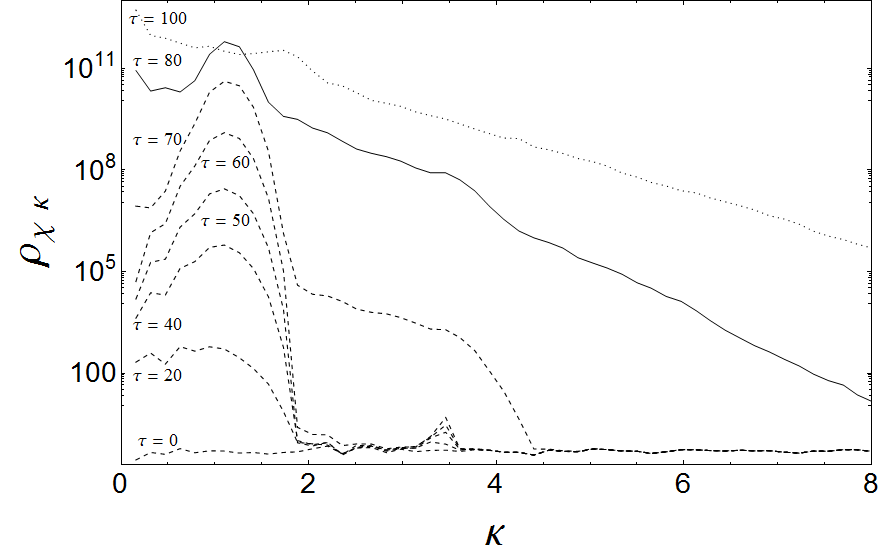}
                        \caption{\label{fig:spectrum_120_0}}
	\end{subfigure}%
	\begin{subfigure}[b]{0.5\textwidth}
		\centering
		\includegraphics[width=\textwidth]{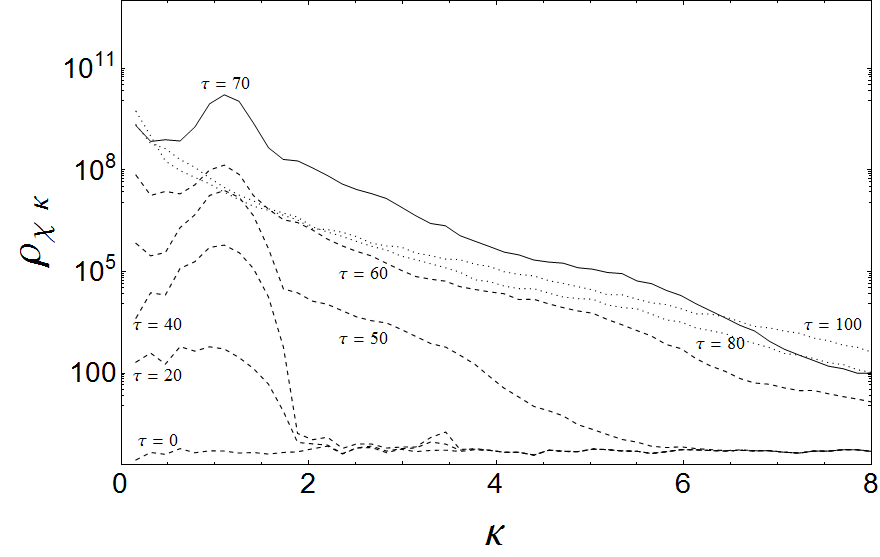}
	            \caption{\label{fig:spectrum_120_8}}
	\end{subfigure}
	\caption{\label{fig:preheat_120} Evolution of preheating for $q = 120$. (a) Energy density $\rho_{\chi}$ as a function of time for $\lambda_{\chi}=0$. (b) Energy density $\rho_{\chi}$ as a function of time for $\lambda_{\chi}=10^{-8}$. (c) Energy density of $\phi$ and $\chi$, as well as energy density in the interaction term, for $\lambda_{\chi}=0$. (d) Same as (c), but for $\lambda_{\chi} = 10^{-8}$. The spatially averaged quantity $q\langle \frac{1}{4}\lambda_{\chi}\chi^4\rangle$ is also shown. (e) The spectrum in $\chi$ at several times of interest, for $\lambda_{\chi}=0$. The solid line corresponds to approximately the time when the exponential growth ends. (f) Same as (e), for $\lambda_{\chi}=10^{-8}$.}
\end{center}
\end{figure}


These can be studied by lattice simulation, and we have used the C++ code LATTICEEASY \cite{Felder:2000hq} in order to simulate the evolution of these interacting scalar fields in an expanding universe. 
\fref{fig:preheat_120} shows results for $q = 120$. This is a useful example since \cite{Dufaux:2007pt} presents detailed results for preheating and gravitational wave production for $q=120$ in the absence of a self-coupling. \fref{fig:egychi_120_0} shows the spatially-averaged energy density $\rho_{\chi} \equiv \langle \frac{1}{2}\dot{\chi}^2 + \frac{1}{2a^2}(\partial_j\chi)^2 + \frac{1}{4}\lambda_{\chi}\chi^4 \rangle$ as a function of time. \fref{fig:egydens_120_0} shows for $\lambda_{\chi}=0$ the sum of the spatially-averaged energy densities $\rho_{\phi} \equiv \langle \frac{1}{2}\dot{\phi}^2 + \frac{1}{2a^2}(\partial_j\phi)^2 + \frac{1}{4}\lambda\phi^4 \rangle$ and $\rho_{\chi}$, as well as the energy density only in the interaction term, $\rho_{\rm int} \equiv \langle \frac{1}{2}g^2\phi^2\chi^2 \rangle$. \fref{fig:spectrum_120_0} shows the spectrum in $\chi$ for the same choice of parameters. The spectrum is shown at several times, and the solid line corresponds to approximately the time when the exponential growth ends.

We can understand how preheating progresses by observing that the transfer of energy between $\chi$ and $\phi$, and among different modes $\chi_{\kappa}$ and $\phi_{\kappa}$, occurs in the following distinct stages. First, oscillations of the homogeneous $\phi$ excite modes of $\chi$ centered around some $\kappa = \kappa_{\ast}$, and the initially small inhomogeneities of $\chi$ become large. There is some backreaction onto $\phi$, whereby the $\frac{g^2}{2}\phi^2\chi^2$ interaction term broadly excites modes $\phi_{\kappa}$ up through $\approx 2 \kappa_{\ast}$, and inhomogeneities in $\phi$ begin to grow.

The second stage occurs once $q^{1/2}\rho_{\rm int} \approx \rho_{\phi}+\rho_{\chi}$. This is a useful, approximate numerical result, that is essentially the same as Eq. 6 in \cite{Khlebnikov:1996zt}. Then $\chi_{\kappa_{\ast}}$ efficiently rescatters, i.e. interacts with other modes, and its exponential growth ends. The total energy in $\chi$ continues to grow a bit until $\rho_{\chi}\approx\rho_{\phi}$. This is evident in \fref{fig:egychi_120_0}. Large field inhomogeneities break up and the spectrum broadens towards larger $k$. This broad spectrum where energy density becomes approximately evenly distributed among modes is evident in \fref{fig:spectrum_120_0}. This figure indicates the spectrum at the time when the exponential growth ends with a solid curve. Spectra before this time are indicated by dashed curves, and spectra after this time are indicated by dotted curves. This stage is discussed and examples of field configurations are shown in \cite{Felder:2006cc}. Some work has also examined the final, so-called ``turbulent thermalization" stage in detail \cite{Micha:2002ey,Micha:2004bv}.

We now consider the case of nonzero $\lambda_{\chi}$. This has been studied to some extent in \cite{Prokopec:1996rr,Allahverdi:1996xc,Frolov:2008hy}, and here we find results consistent with theirs. \fref{fig:egychi_120_8} shows $\rho_{\chi}$ as a function of time. The resonance ends earlier in comparison with the $\lambda_{\chi}=0$ situation of \fref{fig:egychi_120_0}. \fref{fig:egydens_120_8} shows $\rho_{\phi}+\rho_{\chi}$, $\rho_{\rm int}$ and $\langle \frac{1}{4}\lambda_{\chi}\chi^4 \rangle$ for $\lambda_{\chi} = 10^{-8}$. \fref{fig:spectrum_120_8} shows the spectrum in $\chi$ at several times of interest, and the solid line again corresponds to approximately the time when the exponential growth ends. Here the end of this stage still corresponds to a large mixing between modes, but in this case it is the quartic self-interaction that is significant.

In general, we find from numerical simulation that when $\lambda_{\chi}$ becomes significantly larger than $g^2$, the resonance terminates earlier than for the $\lambda_{\chi}=0$ case, i.e. for any $\lambda_{\chi} > \lambda_{\chi}^{\ast} \sim g^2$. In terms of energy transfer, when $q^{1/2}\left\langle \frac{1}{4}\lambda_{\chi}\chi^4 \right\rangle \approx \rho_{\rm int}$, the resonance ends. This is analogous to the condition we described for $\lambda_{\chi}=0$, and will be useful. Depending on the size of $\lambda_{\chi}$, this may occur before or after the relation $q^{1/2}\rho_{\rm int} \approx \rho_{\phi}+\rho_{\chi}$ becomes true. To summarize, the resonant stage of preheating ends by the following condition:
\begin{align}
	\left( \rho_{\phi} + \rho_{\chi} \right) \approx q^{1/2} \rho_{\rm int} & \, \, \, \, \, \, \, \, \, \, \, {\rm for \, \, }\lambda_{\chi}<\lambda_{\chi}^{\ast}, \label{eq:end_1}\\
	\rho_{\rm int} \approx q^{1/2}\left\langle \frac{1}{4}\lambda_{\chi}\chi^4 \right\rangle & \, \, \, \, \, \, \, \, \, \, \, {\rm for \, \, }\lambda_{\chi}>\lambda_{\chi}^{\ast}.  \label{eq:end_2}
\end{align}
The powers of $1/2$ are approximate -- when comparing the size of the oscillating energy densities, as in \fref{fig:egydens_120_0} and \fref{fig:egydens_120_8} for example, there is some ambiguity in determining exactly what the value of the energy is when the resonance ends. We can estimate the value $\lambda_{\chi}^{\ast}$ where the condition \eref{eq:end_2} becomes more important than \eref{eq:end_1} in terms of an energy argument. For small enough $\lambda_{\chi}$, we will have $\langle \frac{1}{4}\lambda_{\chi}\chi^4\rangle \ll \rho_{\phi}+\rho_{\chi}$, so the self-interaction will not play a role in ending the resonance. This will no longer be true once 
\begin{align}
q^{1/2}\left\langle \frac{1}{4}\lambda_{\chi}\chi^4\right\rangle \sim \rho_{\phi}+\rho_{\chi}.
\end{align}
This can be related to the value of $\chi$ when the resonance ends by observing that, around this critical value $\lambda_{\chi}^{\ast}$ where behavior transitions from \eref{eq:end_1} to \eref{eq:end_2}, we will also have
\begin{align}
\rho_{\phi} + \rho_{\chi} \sim q^{1/2}\left\langle \frac{1}{2}g^2\phi_{\rm end}^2\chi_{\rm end}^2 \right\rangle
\end{align}
so that 
\begin{align}\label{eq:end}
\frac{1}{4}\lambda_{\chi}\langle\chi_{\rm end}^4\rangle \sim \frac{1}{2}g^2\langle\phi_{\rm end}^2\chi_{\rm end}^2\rangle
\end{align}
For $( \langle \phi^2\chi^2 \rangle / \langle \chi^4 \rangle )_{\rm end} \sim \mathcal{O}(1)$ this means that 
\begin{align}\label{eq:lambda_star}
\lambda_{\chi}^{\ast} \sim g^2.
\end{align}
This agrees with numerical results showing that the maximum energy density begins to decrease dramatically with increasing $\lambda_{\chi}$ around this value. For example, $q=120$ will give $\lambda_{\chi}^{\ast} \sim 120\times 10^{-13} \sim 10^{-11}$. We check this by defining for each $\lambda_{\chi}$ the quantity $\rho_{\chi}^{\rm max}$ as the time average over several oscillations once $\rho_{\chi}$ has stopped increasing with time. \fref{fig:compare_gw_120} shows that around $\lambda_{\chi}^{\ast} \approx 10^{-11}$, $\rho_{\chi}^{\rm max}$ begins to decrease as $\lambda_{\chi}^{-1}$. We now seek to quantify the effect that this has on gravitational wave production.

\section{Gravitational Wave Spectrum}\label{sec:gw_spectrum}

The metric perturbation $h_{ij}$ defined in \eref{eq:metric} can be rescaled as $\overline{h}_{ij} \equiv a h_{ij}$. Neglecting a term that goes as $a''/a \sim (aH)^2$ \cite{Dufaux:2007pt}, the equation of motion is 
\begin{align}\label{eq:h_eom}
	\overline{h}_{ij}'' - \nabla^2\overline{h}_{ij} & = 16\pi G a^3 \Pi_{ij}^{\rm TT}
\end{align}
where $G$ is Newton's constant and $\Pi_{ij}^{\rm TT}$ is the transverse traceless projection of the anisotropic stress: 
\begin{align}\label{eq:stress}
	\Pi_{ij} = a^{-2}\left( T_{ij} -\langle p \rangle g_{ij} \right).
\end{align}
The second term in \eref{eq:stress} will be neglected since $g_{ij}$ is the sum of a homogeneous, isotropic part whose transverse traceless projection is zero, and a perturbation that is higher order in $G$. The Fourier Transform of \eref{eq:h_eom} is 
\begin{align}
	\overline{h}_{ij}''(\vec{k}) + k^2\overline{h}_{ij}(\vec{k}) & = 16\pi G a^3 \Pi_{ij}^{\rm TT}(\vec{k})
\end{align}

We consider $\Pi_{ij}^{\rm TT}$ to be a source acting continuously during the time interval $\eta_0 < \eta < \eta_f$, solve \eref{eq:h_eom} using Green's functions, and use this solution to find the energy density of the tensor perturbation. As shown in \cite{Dufaux:2007pt}, the result of this procedure is
\begin{align}\label{eq:gw_spec}
	\frac{d \rho_{\rm gw}}{d \ln k}(\eta > \eta_f) & = \frac{S_k}{a^4(\eta)}
\end{align}
where $S_k$ is defined by 
\begin{align}\label{eq:sk_def}
	S_k & = \frac{4\pi G k^3}{V} \int d\Omega \sum_{i,j}\left( \left|\int_{\eta_i}^{\eta_f}d\eta'\cos(k\eta')a(\eta')T_{ij}^{\rm TT}(\eta',\vec{k})\right|^2 + \left|\int_{\eta_i}^{\eta_f}d\eta'\sin(k\eta')a(\eta')T_{ij}^{\rm TT}(\eta',\vec{k})\right|^2 \right)
\end{align}
where $V$ is the volume of the box considered and $\int d\Omega$ is an integral over directions in $k$ space.\footnote{Our physical results are independent of box size, as we use a numerical Fourier Transform that takes this into account. This is described in the LATTICEEASY documentation.} $S_k$ only depends on the dynamics occurring during gravitational wave generation, and the TT part of the energy-momentum tensor is defined in terms of projection operators by 
\begin{align}
	T_{ij}^{\rm TT}(\eta,\vec{k}) & = \left( P_{il}(\hat{k})P_{jm}(\hat{k}) - \frac{1}{2}P_{ij}(\hat{k})P_{lm}(\hat{k}) \right) T_{lm}(\eta,\vec{k}) \\
	P_{ij}(\hat{k}) & = \delta_{ij} - \hat{k}_i\hat{k}_j
\end{align}

We obtain the spectrum of gravitational waves numerically using the LATTICEEASY code mentioned above, modified to in order to compute \eref{eq:gw_spec} as described above. We will give results in terms of $\Omega_{\rm gw} = \rho_{\rm gw} / \rho_{\rm total}$, at the ``time of production" defined as approximately the time when energy in gravitational waves stops increasing noticeably. This is very well approximated by the value at the end of the simulation at $\tau=250$, and denote with a subscript ``p" quantities evaluated at this time. The relation between the results we give and their present values depends somewhat on the equation of state throughout reheating, but previous works have established that in $\lambda\phi^4$ preheating, the equation of state very rapidly becomes that of radiation, so that the energy density in gravitational waves will be \cite{Dufaux:2007pt}
\begin{align}\label{eq:gw_amp_today}
h^2\Omega_{\rm gw} & = \left(\frac{S_k}{a^4\rho}\right)_{p}\left(\frac{g_{0}}{g_{\ast}}\right)^{1/3}h^2\Omega_{\rm rad} \nonumber \\
	& = \left( 9.3 \times 10^{-6} \right)\left(\Omega_{\rm gw}\right)_{p}
\end{align}
where $h^2\Omega_{\rm rad} = 4.3\times 10^{-5}$, $g_{\ast}/g_0 \approx 100$. Similarly, frequencies today are related to comoving wave numbers at the time of preheating by 
\begin{align}\label{eq:gw_freq_today}
	f = \left(\frac{k}{a\rho^{1/4}}\right)_{p}4\times 10^{10} \, {\rm Hz} \, \sim \, \kappa \times 10^{7} \, {\rm Hz}
\end{align}
where in the last step we have taken $(a^4\rho)_p \sim \lambda\phi_0^4$ (see e.g. \fref{fig:egychi_120_0}; we begin with $\rho_{\chi} \approx 0$ and $\rho_{\phi} \approx 2\times \frac{1}{4}\lambda\phi_0^4$ and throughout the simulation the quantity $a^4(\rho_{\phi} + \rho_{\chi}) \approx$ constant).


\begin{figure}[t]
\begin{center}
	\begin{subfigure}[b]{0.5\textwidth}
		\centering
		\includegraphics[width=\textwidth]{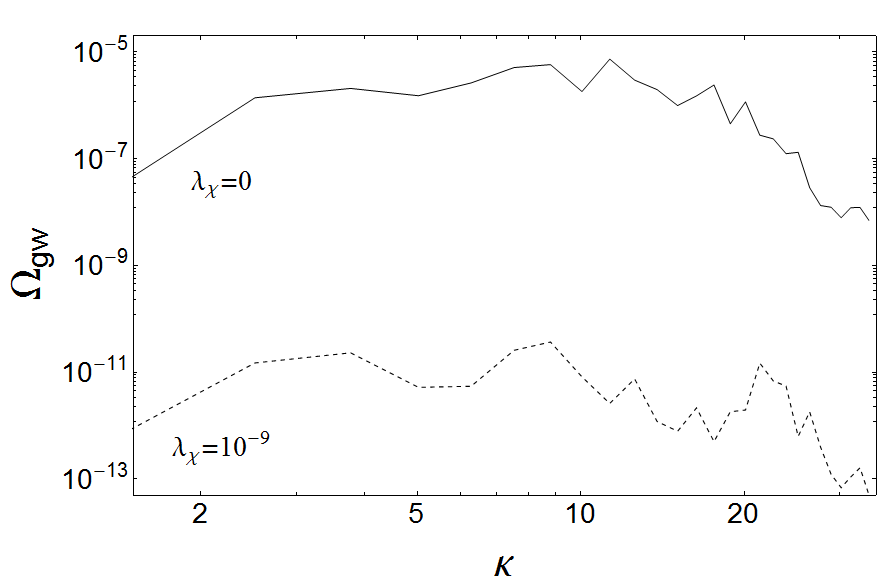}
                        \caption{\label{fig:gwavespec_0and9_120}}
	\end{subfigure}%
	\begin{subfigure}[b]{0.5\textwidth}
		\centering
		\includegraphics[width=\textwidth]{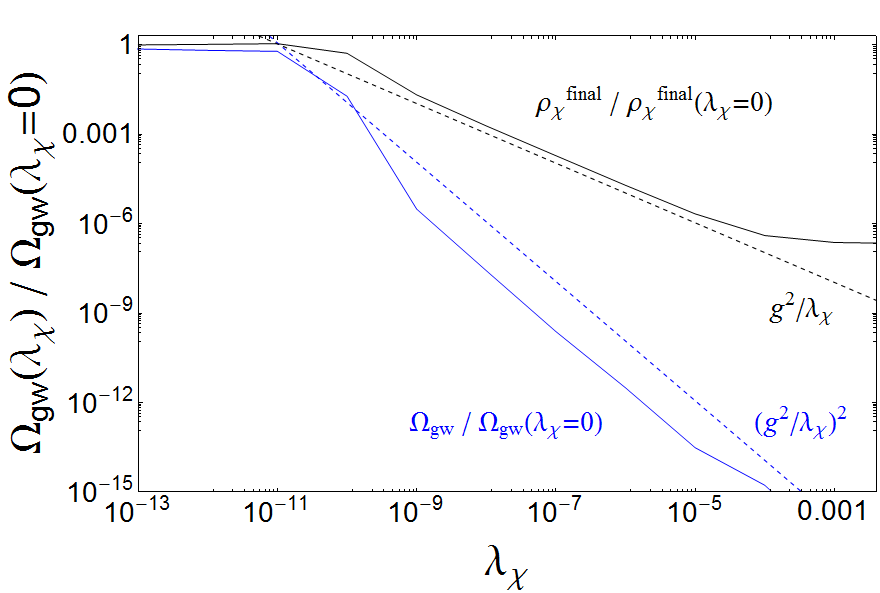}
                        \caption{\label{fig:compare_gw_120}}
	\end{subfigure}
	\begin{subfigure}[b]{0.5\textwidth}
		\centering
		\includegraphics[width=\textwidth]{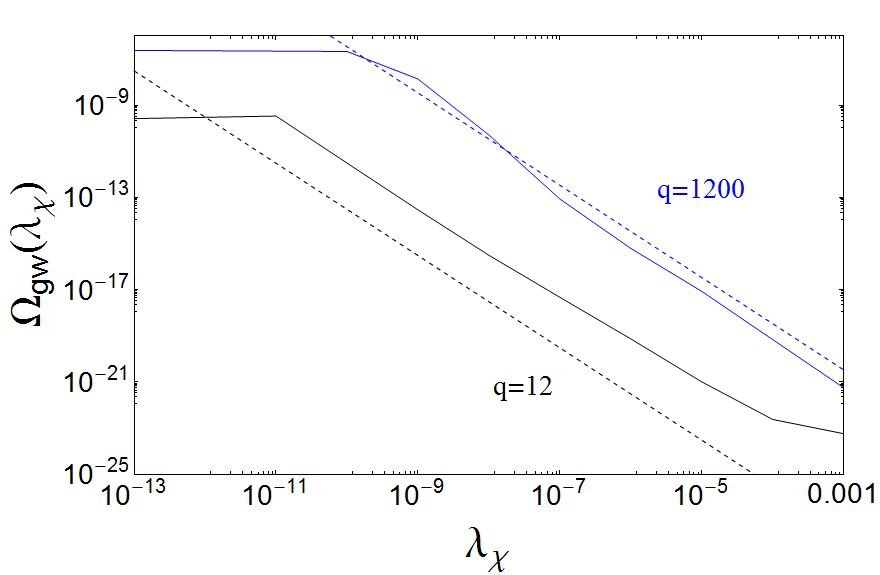}
                        \caption{\label{fig:compare_gw_12and1200}}
	\end{subfigure}%
	\begin{subfigure}[b]{0.5\textwidth}
		\centering
		\includegraphics[width=\textwidth]{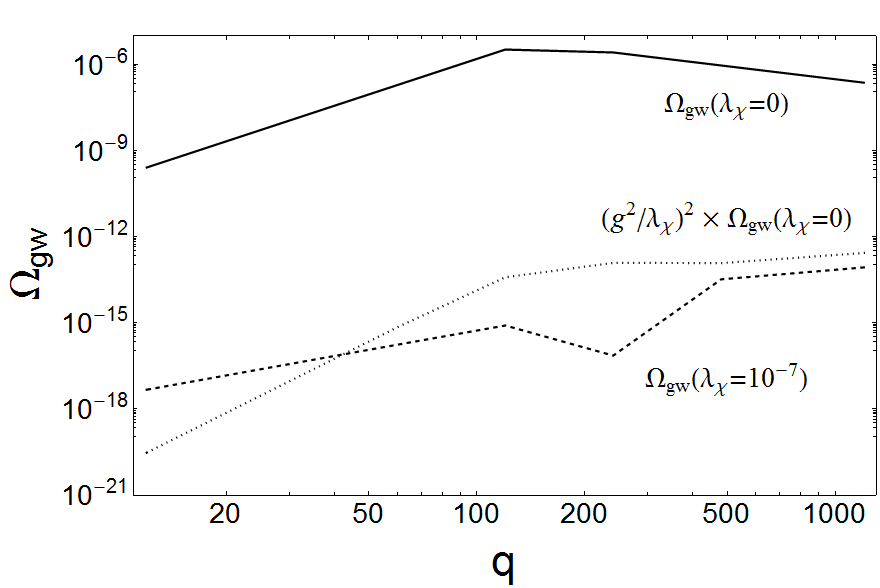}
	\caption{\label{fig:q_dependence}}
	\end{subfigure}

	\caption{\label{fig:gw_spectra} Peak of gravitational wave energy density spectrum, defined in \eref{eq:gw_spec}, as a fraction of total energy density at end of preheating stage. (a) Spectrum for $q=120$ and two choices of self-coupling $\lambda_{\chi}$. (b) Amplitude of peak of GW spectrum, and final average value for $\rho_{\chi}$ after preheating ends, for $q=120$ and as a function of $\lambda_{\chi}$. These quantities are presented as fractions of their value in the $\lambda_{\chi}=0$ case. For comparison, dashed curves are also shown for the scaling behavior \eref{eq:rho_scaling} and \eref{eq:gw_scaling}. (c) Amplitude of peak of GW spectrum, $\Omega_{\rm gw}^{\ast}$, as a function of $\lambda_{\chi}$ for $q=12$ and $q=1200$, compared with \eref{eq:gw_scaling}. (d) Value of $\Omega_{\rm gw}$ as a function of the resonance parameter, $q$, for $\lambda_{\chi}=0$ and $\lambda_{\chi}=10^{-7}$, and the prediction \eref{eq:gw_scaling} applied to the latter case.}
\end{center}
\end{figure}


\fref{fig:gwavespec_0and9_120} shows the spectrum obtained in the case $q=120$, for $\lambda_{\chi} = 0$ and $\lambda_{\chi} = 10^{-9}$. The decrease in the energy produced in gravitational waves is evident from this, and \fref{fig:compare_gw_120} shows how this depends on $\lambda_{\chi}$, as a fraction of the peak energy density when $\lambda_{\chi} = 0$. The solid lines in \fref{fig:compare_gw_12and1200} show $\Omega_{\rm gw}$ for the cases $q=12$ and $q=1200$. Evidently the effect of $\lambda_{\chi}$ is to end the resonance early and suppress gravitational wave production. Once preheating ends, the additional contribution of inhomogeneities to the gravitational wave spectrum is negligible \cite{Dufaux:2007pt}.

To estimate how this effect depends on the model parameters $q$ and $\lambda_{\chi}$, we note that $\Omega_{\rm gw} \sim (T_{ij}^{\rm TT})^2 \sim (\partial_i \chi)^4$. The energy density $\Omega_{\rm gw}$ is dominated by the most recently produced part of the spectrum before the resonance ends (this is particularly clear in Fig. 8 of \cite{Dufaux:2007pt}), so for the purposes of this estimate we will ask how the maximum amplitude of $\chi$ depends on $q$ and $\lambda_{\chi}$. We have seen that $\chi$ grows until the condition \eref{eq:end_2}, $\frac{1}{2}q\lambda\langle\phi^2\chi^2\rangle \sim \frac{1}{4}\lambda_{\chi}\langle\chi^4\rangle$, is satisfied. (Also, comparison of \fref{fig:egychi_120_0} with \fref{fig:egychi_120_8} shows this since $\rho_{\chi} \sim (\partial_{i}\chi)^2$.) This suggests a parametric scaling
\begin{align}\label{eq:rho_scaling}
	\chi_{\rm end}^2 \propto q \lambda / \lambda_{\chi} = g^2 / \lambda_{\chi}
\end{align}
Then the expectation that $\Omega_{\rm gw} \sim (\chi_{\rm end}^2)^2$ becomes
\begin{align}\label{eq:gw_scaling}
	\Omega_{\rm gw} \propto \left(g^2 / \lambda_{\chi}\right)^2.
\end{align}

Our numerical results confirm this relation as shown in \fref{fig:compare_gw_120} and \fref{fig:compare_gw_12and1200}. For $\lambda_{\chi} < \lambda_{\chi}^{\ast}$, the peak energy in gravitational waves decreases only very slightly with increasing $\lambda_{\chi}$, as the self-interaction term plays a small role in mixing modes and damping inhomogeneities. Once $\lambda_{\chi} > \lambda_{\chi}^{\ast}$, the energy density in gravitational waves scales in the manner given by \eref{eq:gw_scaling}. For $\lambda_{\chi} \sim 10^{-2}$, we see that $\rho_{\chi}$ and $\Omega_{\rm gw}$ no longer decrease significantly with increasing $\lambda_{\chi}$. This is simply because the unstable resonance never begins, and the quartic self-interaction can no longer dramatically decrease $\Omega_{\rm gw}$ by ending the resonance earlier. \fref{fig:q_dependence} shows how the value of $\Omega_{\rm gw}$ at the time of production depends on the resonance parameter, $q$, for both $\lambda_{\chi}=0$ and $\lambda_{\chi}=10^{-7}$. In the latter case, we also show the prediction of the scaling relation \eref{eq:gw_scaling}.

\section{Generality}\label{sec:generality}

So far, we have examined results in the context of $\lambda\phi^4$ chaotic inflation, with the self-coupling $\lambda$ and the initial condition of the inflaton field identical to a previous work that thoroughly investigated the dynamics of gravitational wave production during preheating \cite{Dufaux:2007pt}. This allows the results of the previous sections to be directly compared with that work. However, observational data indicates that the $\lambda\phi^4$ chaotic potential is not favored \cite{Ade:2015oja}, so an important question is the generality of the results we have quoted above. In this section we will address this question in two ways, before pointing out interesting directions for future work. We will consider massive ($m^2\phi^2$) inflation, another standard example in which preheating is studied, and we will also consider a range of initial conditions for $\phi$ within both the $\lambda\phi^4$ and $m^2\phi^2$ cases.

Specifically, this means that we will begin the numerical situation -- corresponding to the end of inflation, with the inflaton's energy about evenly split between kinetic and potential -- with the inflaton field at various lower points on its potential than in the original case. Here, we are not primarily concerned with representing a complete model of inflation, but rather are studying how preheating and gravitational wave production proceed within a potential that is quadratic or quartic about the minimum, without regard to the model's behavior at higher (inflationary) field values. In this spirit, we also study the $m^2\phi^2$ case with a few choices of $m_{\phi}$. It is worth pointing out that not all inflationary models end with oscillations of the field responsible for inflation about its zero; see for example the Abelian Higgs and Higgs-dilaton models \cite{Dufaux:2010cf,GarciaBellido:2011de}.


\begin{figure}[t]
		\includegraphics[width=0.5\textwidth]{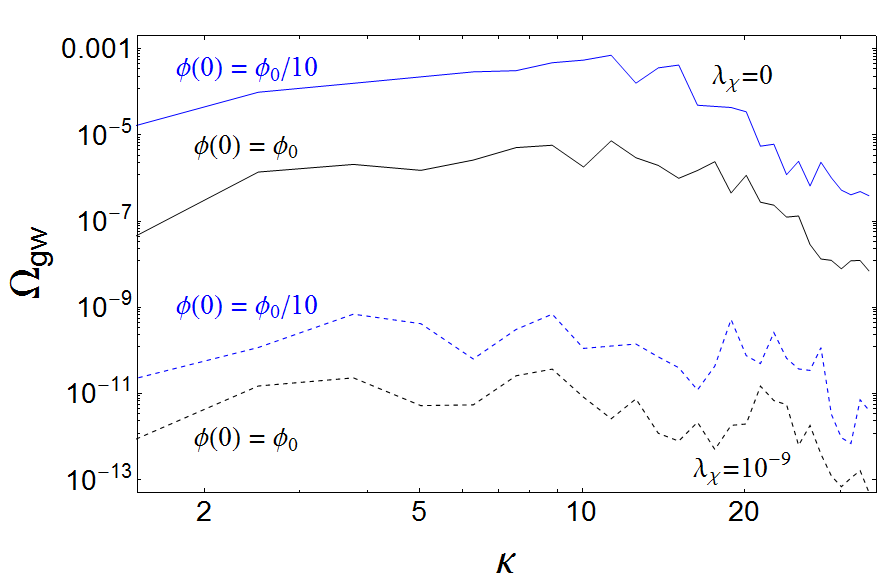}
             	\caption{\label{fig:gwavespec_phi4_newic} Peak of gravitational wave energy density spectrum, defined in \eref{eq:gw_spec}, as a fraction of total energy density at end of preheating stage. Spectrum for $q=120$ and two choices of self-coupling $\lambda_{\chi}$. The curves labeled $\phi(0) = \phi_0$ are identical to those shown in \fref{fig:gwavespec_0and9_120}, corresponding to the original initial condition for the inflaton field. The curves labeled $\phi(0) = \phi_0/10$ correspond to starting the inflaton a factor of 10 lower on the potential, as described in the text. The magnitude of the gravitational wave spectrum is changed, but the effect of turning on $\lambda_{\chi}$ is the same.}

\end{figure}

\begin{figure}[t]
\begin{center}
	\begin{subfigure}[b]{0.5\textwidth}
		\centering
		\includegraphics[width=\textwidth]{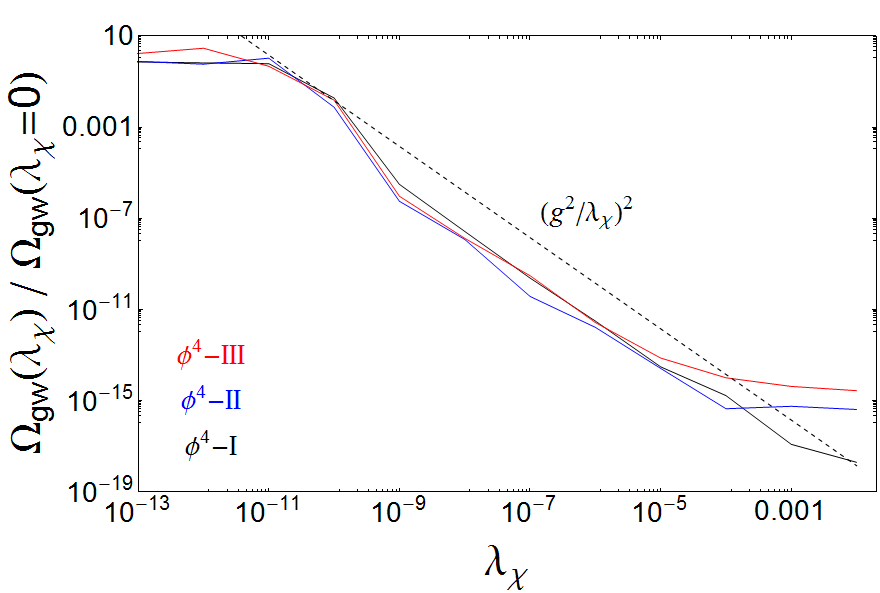}
                        \caption{\label{fig:compare_gw_phi4_newic}}
	\end{subfigure}%
	\begin{subfigure}[b]{0.5\textwidth}
		\centering
		\includegraphics[width=\textwidth]{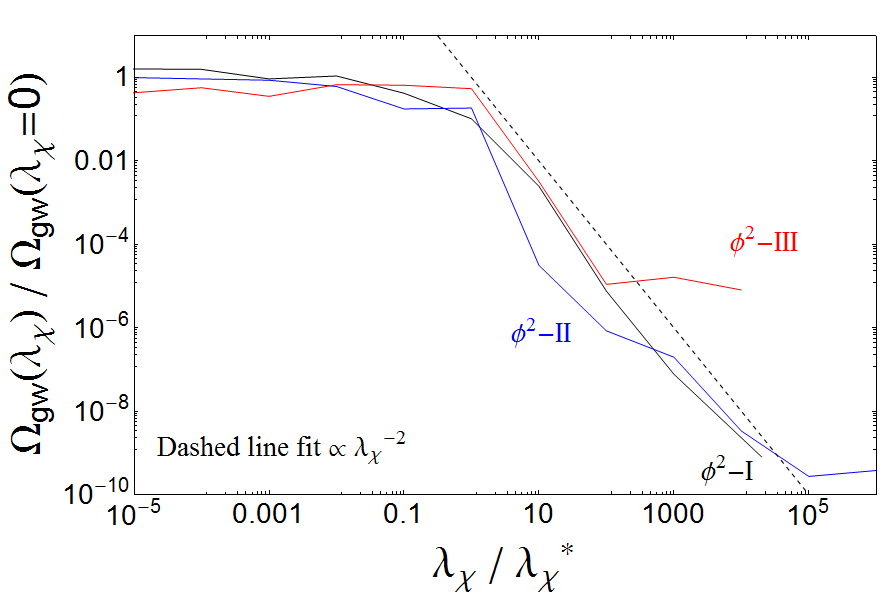}
                        \caption{\label{fig:compare_gw_m2phi2_newic}}
	\end{subfigure}

	\caption{\label{fig:gwcomp_gen} Peak of gravitational wave energy density spectrum, defined in \eref{eq:gw_spec}, as a fraction of total energy density at end of preheating stage and normalized to the value when $\lambda_{\chi}=0$. Here we show several typical examples where the initial condition and/or parameters of the model are varied, as described in the text. As in \sref{sec:gw_spectrum}, there is some value $\lambda_{\chi}^{\ast}$ above which the peak of the gravitational wave spectrum decreases as $\lambda_{\chi}^{-2}$. Once $\lambda_{\chi}$ is large enough, the preheating resonance never starts and there is no further suppression with increasing $\lambda_{\chi}$, an effect also seen in \sref{sec:gw_spectrum}. (a) Varying initial conditions for $\lambda\phi^4$ inflaton potential. (b) Varying initial conditions and mass parameter for $m^2\phi^2$ inflaton potential. For ease of comparison, this result is given as a function of $\lambda_{\chi} / \lambda_{\ast}$, where $\lambda_{\ast}^{\rm I} = 10^{-6}$, $\lambda_{\ast}^{\rm II} = 10^{-9}$, $\lambda_{\ast}^{\rm III} = 10^{-7}$.}
\end{center}
\end{figure}

For every situation we have tried, the same approximate scaling behavior of reduced gravitational wave production with increased self-interaction $\lambda_{\chi}$ holds. In particular, we display some typical results in \fref{fig:gwavespec_phi4_newic} and \fref{fig:gwcomp_gen}. For the case of $\lambda\phi^4$ inflation, with $q = 120$ and $\phi(0) = \phi_0 \equiv 0.342 \, M_{\rm Pl}$, we plot the gravitational wave spectrum in \fref{fig:gwavespec_phi4_newic} and the scaling behavior with $\lambda_{\chi}$ in \fref{fig:compare_gw_phi4_newic}. These results were presented in \sref{sec:gw_spectrum}, and they are provided again for direct comparison with alternative scenarios. We label this choice of parameters as $\phi^4-{\rm I}$. We also show results for $q = 120$ and $\phi(0) = \phi_0 / 10$, referred to as $\phi^4-{\rm II}$, as well as $q = 120$ and $\phi(0) = \phi_0 / 100$, referred to as $\phi^4-{\rm III}$.


\fref{fig:gwavespec_phi4_newic} compares the gravitational wave spectra of the $\phi^4-{\rm I}$ and $\phi^4-{\rm II}$ parameter choices, for both $\lambda_{\chi} = 0$ and $\lambda_{\chi} = 10^{-9}$. In both cases, evidently, there is a significant reduction in gravitational wave production that accompanies an increase in $\lambda_{\chi}$, despite the difference in overall amplitude of the spectrum. In \fref{fig:compare_gw_phi4_newic}, we show how this reduction depends on $\lambda_{\chi}$ for each of the parameter choices $\phi^4-{\rm I}$, $\phi^4-{\rm II}$, $\phi^4-{\rm III}$. We find the same scaling behavior as before: there is a $\lambda_{\chi}^{\ast}$ above which gravitational wave production is suppressed by a factor of $(g^2 / \lambda_{\chi} )^2$.

In the case of massive inflation, we replace \eref{eq:potential} with the potential 
\begin{align}\label{eq:m2p2_potential}
	V = \frac{1}{2}m_{\phi}^2\phi^2 + \frac{1}{4}\lambda_{\chi}\chi^4 + \frac{1}{2}g^2\phi^2\chi^2
\end{align}
i.e. the light field $\chi$ has the same potential and interactions with the inflaton as it did previously, but the inflaton potential is quadratic rather than quartic. In this case we find that, as above, there is some $\lambda_{\chi}^{\ast}$ such that for $\lambda_{\chi} > \lambda_{\chi}^{\ast}$, gravitational wave production tends to be suppressed by $\lambda_{\chi}^{-2}$. We again plot three typical examples. We refer to $q \equiv g^2\phi(0)^2 / 4m_{\phi}^2 = 60$, $\phi(0) = 0.1 \, M_{\rm Pl}$, $m_{\phi} = 10^{-6}M_{\rm Pl}$ as $\phi^2-{\rm I}$. We refer to $q = 15$, $\phi(0) = 0.01$, $m_{\phi} = 10^{-9}$ as $\phi^2-{\rm II}$. We refer to $q = 15$, $\phi(0) = 0.001$, $m_{\phi} = 10^{-9}$ as $\phi^2-{\rm III}$. \fref{fig:compare_gw_m2phi2_newic} shows how the gravitational wave spectra in these cases scale with $\lambda_{\chi}$. For ease of comparison with the scaling relation $\lambda_{\chi}^{-2}$ we plot the results as a function of $\lambda_{\chi}/\lambda_{\chi}^{\ast}$, where $\lambda_{\chi}^{\ast} = 10^{-6}, \, 10^{-9}, \, 10^{-7}$ for $\phi^2-{\rm I}$, $\phi^2-{\rm II}$, $\phi^2-{\rm III}$ respectively. As before, $\lambda_{\chi}^{-2}$ fits well (until $\lambda_{\chi}$ becomes large enough that preheating no longer starts, so that increasing $\lambda_{\chi}$ won't further decrease the gravitational wave production).

The numerical computations involved make it impractical to check here every imaginable situation of interest to verify this relation. We have shown that gravitational wave production from preheating in potentials with minimum at zero can be extremely sensitive to the value of the light field's self-coupling term, and that result is not exclusive to one particular model or choice of parameters. Therefore, an important goal of future work will be to fully characterize this effect in other realistic models, and better understand the implications for observability.

\section{Discussion and Conclusions}\label{sec:discussion}

In this work we have studied the effect of a nonzero self-interaction on gravitational wave production during preheating of a scalar field. Previous work has considered the dynamics of preheating for a light, self-interacting scalar, as well as gravitational wave production by preheating of a non-self-interacting scalar. This work is an extension of these results, and in particular shows that the spectrum of gravitational waves that survive until today is very sensitive to the light scalar's self-interaction. Our main result within the $\lambda\phi^4$ model is that for self-coupling $\lambda_{\chi} \gtrsim g^2$, the preheating resonance is terminated early, and the gravitational wave spectrum is significantly reduced: 
\begin{align}\label{eq:result1}
	\Omega_{\rm gw} \approx \left(\frac{g^2}{\lambda_{\chi}}\right)^2 \Omega_{\rm gw}^{(\lambda_{\chi}=0)} \, \, \, \, \, \, \, \, {\rm for} \, \, \, \, \, \, \, \, \lambda_{\chi} \gtrsim g^2.
\end{align}

We have also begun to address the question of generality of this result, as discussed in \sref{sec:generality}. For various choices of the inflaton's initial condition in the $\lambda\phi^4$ model, we have seen that \eref{eq:result1} holds. Additionally, for an $m^2\phi^2$ inflationary potential, the result that the gravitational waves are suppressed as $\lambda_{\chi}^{-2}$ is shown, for several parameter choices. While this suggests generality to inflation models with potentials quadratic or quartic about a minimum at zero, an important question for future work is to study the effect of the light field's interactions in other preheating models that have been shown to predict gravitational waves. As our work shows, predictions that neglect such interactions - even if they are extremely small - may not necessarily be accurate.

It is easy to imagine that in a realistic preheating scenario, decay products will have their own self-interactions or further interactions with other fields, that will end the resonance early.
Recently, another paper studied the effect of interactions of $\chi$ with further light degrees of freedom, as well as self-interactions in the context of a curvaton decaying to Higgs \cite{Lerner:2015uca}. Although the model is not identical to ours, it also found that self-interactions can be important in terminating the resonance early. Furthermore, they found that interaction with the additional light scalars, as characterized by the contribution to a thermal term, has the ability to significantly affect the resonance and either end it early or prevent it from occurring at all. 
They did not consider gravitational wave production, but following the argument given here in \sref{sec:gw_spectrum} it is reasonable to expect that this early termination of the resonance can further reduce any production of gravitational waves. Analyses of other scenarios have shown that preheating can be sensitive to nonlinear interaction terms of decay products \cite{Enqvist:2014tta}, or other nonperturbative effects motivated by new physics above the TeV scale \cite{Kusenko:2008zm,Kusenko:2009cv,Chiba:2009zu,Zhou:2013tsa,Zhou:2015yfa}. Another interesting goal for future work would be to incorporate the effects of interactions such as those studied in this paper into a more general framework for obtaining order-of-magnitude estimates of gravitational wave production, as in \cite{Giblin:2014dea}. Although current constraints on MHz gravitational wave backgrounds are not sensitive to these processes \cite{Akutsu:2008qv}, this could be very useful in evaluating the potential for observability in future experiments.

One interesting possibility is that reheating occurred through an inflaton-to-Higgs coupling, since the Higgs is a natural candidate to couple to beyond-Standard Model fields \cite{Patt:2006fw,Bhattacharya:2014gva,Kamada:2014ufa,Gross:2015bea}. The running of the Higgs self-coupling is sensitive to any new physics that comes in at high energies, but it has not been directly measured and will be difficult to measure at the LHC. One might hope that since $\lambda_{\rm H}$ runs from 0.13 at the weak scale to zero around $10^{10}$ GeV in the Standard Model \cite{Chetyrkin:2012rz}, the condition $\lambda_{\rm H} \ll 1$ could be satisfied. This would avoid enormous damping of the preheating resonance, and thereby provide a possible cosmological probe of $\lambda_{\rm H}$. The self-coupling remains $\mathcal{O}\left( 10^{-1} \right)$ up to $\sim 10^8$ GeV, though, which suggests that there will not be significant (or any) preheating resonance. However, above this scale the self-coupling decreases and the effective potential reaches a maximum (in the Standard Model -- small changes in input parameters or new physics beyond the Standard Model can significantly affect this; see e.g. \cite{Lebedev:2012zw,Branchina:2013jra,Branchina:2014usa,Branchina:2014rva}). 

The condition \eref{eq:end_2} suggests that a more relevant condition than the self-coupling may be the magnitude of the Higgs potential. The configuration of $\chi$ at the end of inflation (initial configuration for this problem) is certainly sensitive to the potential at large field values, as it corresponds to approximately $\chi_{\rm rms} \sim 10^{12}$ GeV $\sim H_{\rm inf}$.\footnote{The behavior of the Higgs after inflation, when there is no coupling to the inflaton, is discussed in \cite{Enqvist:2013kaa}.} If one takes \eref{eq:end_2} to apply as the condition for whether parametric resonance does or does not occur, then the result could be a resonance pushing Higgs oscillations toward the vacuum instability region.\footnote{There has been much work on the Higgs and vacuum stability, including discussion of the reheat temperature; see e.g. \cite{Espinosa:2007qp,Kobakhidze:2013tn,Degrassi:2012ry} and references therein.} New physics that prevents $\lambda_{\rm H}$ from becoming negative would likely be more than sufficient to prevent a resonance from occuring. These rough estimates also ignore the possibilities of a different running of $\lambda_{\rm H}$ from the new inflaton coupling, as well as thermal effects. We leave the resolution of these questions to future work.

\acknowledgments

I am grateful to Jessica Cook, Emanuela Dimastrogiovanni, Francis Duplessis, Damien Easson, Andrew Long and especially Tanmay Vachaspati for valuable discussions and comments on a draft of this paper. 
I also thank Michael Landry for bringing the references on MHz gravitational wave detection to my attention, and I thank the referee for useful suggestions that have improved this work. 
This work was supported by the U.S. Department of Energy at ASU.  

\bibliography{refs}

\end{document}